\documentclass[twoside]{amsart}
\usepackage{amsmath, amsthm, euscript,epsfig}

\def\Bbb{\mathbb} 
\def\cal{\mathcal} 
\def\Eu{\EuScript}
\def\Ai{\mbox{\rm Ai}} 
  
\def\Re{\mbox{\rm Re}\,} 
 
\def\Im{\mbox{\rm Im}\,} 
\def\varkappa{\mbox{\msbm\char'173}}      
\def\rme{\mbox{\rm e}}
\def\rmi{\mbox{\rm i}}

\theoremstyle{plain} 
\newtheorem{theorem}{Theorem}[section]

\theoremstyle{definition} 

\newtheorem{rem}{Remark}[section]

\numberwithin{equation}{section}

\numberwithin{figure}{section}

\begin{document}

\title{Quasi-linear Stokes phenomenon for 
the Hastings-M{\small c}Leod solution of 
the second Painlev\'e equation}

\author{A.A. Kapaev}
\address{St Petersburg Department of Steklov Mathematical Institute, 
Fontanka 27, St Petersburg 191011, Russia}
\email{kapaev@pdmi.ras.ru}

\begin{abstract}
Using the Riemann-Hilbert approach, we explicitly construct the
asymptotic $\Psi$-function corresponding to the solution
$y\sim\pm\sqrt{-x/2}$ as  $|x|\to\infty$ to the second Painlev\'e
equation $y_{xx}=2y^3+xy-\alpha$.  We precisely describe the
exponentially small jump in the dominant  solution and the coefficient
asymptotics in its power-like expansion.
\end{abstract}

\maketitle
\thispagestyle{empty}
\pagestyle{myheadings} 
\markboth{A.~A.~Kapaev}{Quasi-linear Stokes phenomenon for 
the Hasting-McLeod solution of $P_2$}

\section{Introduction}\label{intro}

The second Painlev\'e equation,
\begin{equation}\label{p2}\tag{$P_2$}
y_{xx}=2y^3+xy-\alpha,\quad \alpha=const,
\end{equation}
was introduced more than a century ago in a classification of the
second order ODEs $y_{xx}=R(x,y,y_x)$ with the Painlev\'e property
\cite{ince}. To this date, equation $P_2$ has found interesting and
important applications in the modern theory of nonlinear waves
\cite{abl_seg, hast_mcleod}, plasma physics \cite{ZKM}, bifurcation
theory \cite{maree}, random matrices and combinatorics \cite{TW, TW2},
theory of semi-classical orthogonal polynomials \cite{BI} and others.

Among its various solutions, we distinguish those with a {\em monotonic} 
asymptotic behavior as $x\to\pm\infty$ (look for complete list of 
asymptotics and connection formulae to transcendent solutions of $P_2$ 
in \cite{kapaev4}). For instance, if $\alpha=0$ and $x\to+\infty$, there 
exists a 1-parameter family of solutions approximated with exponential 
accuracy by decreasing solutions of the Airy equation, $y_{xx}=xy$, see 
\cite{abl_seg},
\begin{equation}\label{Airy_sol}
y\simeq a\mbox{\rm Ai}(x)=
\frac{a}{2\sqrt\pi}x^{-1/4}e^{-\frac{2}{3}x^{3/2}}
\bigl(1+{\cal O}(x^{-3/2})\bigr),\quad 
a=const.
\end{equation}
If in the above asymptotics $0\leq a<1$, then the Painlev\'e function
is bounded and oscillates as $x\to-\infty$,
$$
y\simeq
b(-x)^{-1/4}\sin\bigl(\tfrac{2}{3}(-x)^{3/2}-\tfrac{3}{4}b^2\ln(-x)
+\phi\bigr),
$$
with real constants $b$ and $\phi$ determined by $a$, cf.\ \cite{abl_seg}. 
If $a>1$, then the solution blows up at a finite point. If $a=1$, 
then the solution of $P_2(\alpha=0)$  grows like a square root as 
$x\to-\infty$, $y\simeq\sqrt{-x/2}$, see \cite{hast_mcleod}.

For the first time, the asymptotic as $|x|\to\infty$ behavior of the
Painlev\'e transcendents in the complex domain was studied by Boutroux
\cite{boutroux}. Generically, the Painlev\'e asymptotics within the
sectors $\arg x\in\bigl(\frac{\pi}{3}n,\frac{\pi}{3}(n+1)\bigr)$,
$n\in{\Bbb Z}$  is described by a modulated elliptic sine which
trigonometrically degenerates along the directions  $\arg
x=\frac{\pi}{3}n$.

Besides generic solutions, Boutroux described 1- and 0-parameter reductions 
of the trigonometric asymptotic solutions which admit analytic continuation 
from the ray $\arg x=\frac{\pi}{3}n$ into an adjacent complex sector. As 
$x\to\infty$ in the interior of the relevant complex sector, such solution 
is represented in the leading order by a power sum of $x$ and of 
a {\em trans-series} which is a sum of exponentially small terms,
\begin{equation}\label{gen_struct}
y=\hbox{(power series)}+\hbox{(exponential terms)}.
\end{equation}
Expansions of such kind can be obtained using a conventional
perturbation analysis.

In \cite{kapaev:P1, kapaev:P2}, it was observed that asymptotic solutions 
of the form (\ref{gen_struct}) exhibit a {\em quasi-linear Stokes phenomenon}, 
i.e.\ a discontinuity in a minor term with respect to $\arg x$. The first 
rigorous study of the quasi-linear Stokes phenomenon associated to solutions 
$y\sim\alpha/x$ as $x\to\infty$ for $P_2$ and $y\sim\sqrt{-x/6}$ as 
$x\to\infty$ for $P_1$ is presented in \cite{its_kapaev1, kapaev2}, 
where the reader can find a discussion of other approaches to the same 
problem.

Below, we study a quasi-linear Stokes phenomenon for the second Painlev\'e
transcendent with the monotonic asymptotic behavior $y\sim\sqrt{-x/2}$ as 
$x\to-\infty$. Our main tool is the isomonodromy deformation method 
\cite{JMU, FN, its_nov} in the form of the Riemann-Hilbert problem 
approach via the nonlinear steepest descent method \cite{DZ}.

As in \cite{its_kapaev1, kapaev2}, we pursue a two-fold goal, i.e.\ 
(a)~we bring an exact meaning to the formal expression (\ref{gen_struct}), 
and (b)~we evaluate the asymptotics of the coefficients of the leading 
power series in (\ref{gen_struct}).

\section{Riemann-Hilbert problem for $P_2$}\label{RH_p2}

The inverse problem method in the form of the Riemann-Hilbert (RH)
problem was first applied to an asymptotic study of $P_2$ in \cite{FN}. 
Further study of the RH problem can be found in \cite{DZ, FA, FZ, IFK, 
its_kapaev3}.

According to \cite{FN}, the set of generic Painlev\'e functions is 
parameterized by two of the Stokes multipliers of the associated linear 
system denoted below by the symbols $s_k$, $k=0,1,2$. As it is shown in 
\cite{its_kapaev3}, if $s_0(1+s_0s_1)\neq0$ then the asymptotic solution of 
the RH problem within the sector $\arg x\in(\frac{2\pi}{3},\pi)$ can be 
expressed in terms of elliptic $\theta$-functions which in turn yields 
an elliptic asymptotics to the Painlev\'e function itself. Assuming that 
the condition $s_0=0$ holds true, we arrive to an RH problem which leads 
to a decreasing asymptotics $y\sim\alpha/x$, see \cite{its_kapaev1}.

In the present paper, we first construct an asymptotic solution to the 
relevant RH problem as $|x|\to\infty$, $\arg x\in(\frac{2\pi}{3},\pi)$ 
assuming that 
\begin{equation}\label{Stokes_cond}
1+s_0s_1=0.
\end{equation}

Below, we use the RH problem of Flaschka and Newell \cite{FN} modified as it
is proposed in \cite{its_kapaev1}. This RH problem comes in a standard way 
from a collection of properly normalized solutions of the Lax pair for $P_2$,
\begin{subequations}\label{Lax_pair_p2}
\begin{align}\label{Lax_pair_p2_lambda}
&\frac{\partial\Psi}{\partial\lambda}\Psi^{-1}=
-i\bigl(4\lambda^2+x+2y^2\bigr)\sigma_3
-\bigl(4y\lambda+\frac{\alpha}{\lambda}\bigr)\sigma_2
-2y_x\sigma_1,
\\\label{Lax_pair_p2_x}
&\frac{\partial\Psi}{\partial x}\Psi^{-1}=
-i\lambda\sigma_3-y\sigma_2,
\end{align}
\end{subequations}
where $\sigma_3=\bigl(\begin{smallmatrix}1&0\\0&-1\end{smallmatrix}\bigr)$,
$\sigma_2=\bigl(\begin{smallmatrix}0&-i\\i&0\end{smallmatrix}\bigr)$,  
$\sigma_1=\bigl(\begin{smallmatrix}0&1\\1&0\end{smallmatrix}\bigr)$.

\begin{figure}[hbt]
\begin{center}
\epsfig{file=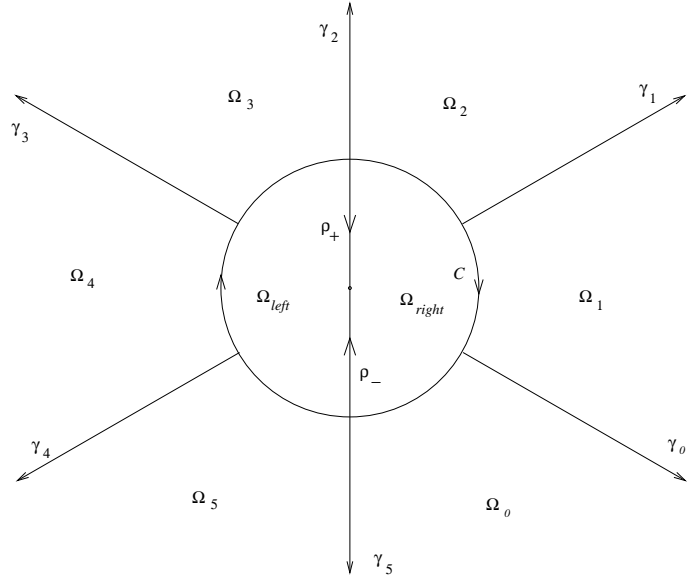}
\caption{A Riemann-Hilbert problem graph for $\Psi(\lambda)$
associated to $P_2$.}
\label{fig1}
\end{center}
\end{figure}

Let us introduce the piece-wise oriented contour 
$\gamma=C\cup\rho_+\cup\rho_-\cup_{k=0}^5\gamma_k$ which is the union 
of the rays 
$\gamma_k=\{\lambda\in{\Bbb C}\colon\
|\lambda|>r,\
\arg\lambda=\frac{\pi}{6}+\frac{\pi}{3}(k-1)\}$, $k=0,1,\dots,5$, oriented
toward infinity, the clock-wise oriented circle 
$C=\{\lambda\in{\Bbb C}\colon\
|\lambda|=r\}$, and of two vertical radiuses
$\rho_+=\{\lambda\in{\Bbb C}\colon\ |\lambda|<r,\
\arg\lambda=\frac{\pi}{2}\}$
and
$\rho_-=\{\lambda\in{\Bbb C}\colon\ |\lambda|<r,\
\arg\lambda=-\frac{\pi}{2}\}$
oriented to the origin. The contour $\gamma$ divides the complex
$\lambda$-plane into 8 regions $\Omega_k$, 
$k\in\{\mbox{\rm left},\mbox{\rm right},0,1,\dots,5\}$.
$\Omega_{\mbox{\rm\tiny left}}$ and $\Omega_{\mbox{\rm\tiny right}}$ are left 
and right halves of the interior of the circle $C$ deprived the radiuses 
$\rho_+$, $\rho_-$. The regions $\Omega_k$, $k=0,1,\dots,5$, are the sectors 
between the rays $\gamma_k$ and $\gamma_{k-1}$ outside the circle, see
Figure~\ref{fig1}. 

Let each of the regions $\Omega_k$, $k=\mbox{\rm right},0,1,2$, be a domain 
for a holomorphic $2\times2$ matrix function $\Psi_k(\lambda)$. Denote the
collection of $\Psi_k(\lambda)$ by $\Psi(\lambda)$,
\begin{equation}\label{Psi_collect}
\Psi(\lambda)\bigr|_{\lambda\in\Omega_k}=\Psi_k(\lambda),\qquad
\Psi(\rme^{\rmi\pi}\lambda)=\sigma_2\Psi(\lambda)\sigma_2.
\end{equation}
Let $\Psi_+(\lambda)$ and $\Psi_-(\lambda)$ be the limits of $\Psi(\lambda)$
on $\gamma$ to the left and to the right, respectively. Let us also introduce 
two matrices 
$\sigma_+=\bigl(\begin{smallmatrix}0&1\\0&0\end{smallmatrix}\bigr)$,
$\sigma_-=\bigl(\begin{smallmatrix}0&0\\1&0\end{smallmatrix}\bigr)$.

The RH problem we talk about is the following one:
\begin{enumerate}
\item
Find a piece-wise holomorphic $2\times2$ matrix function
$\Psi(\lambda)$ such that
\begin{equation}\label{Psi_at_infty}
\Psi(\lambda)e^{\theta\sigma_3}\to I,\qquad
\lambda\to\infty,\qquad
\theta=i\bigl(\tfrac{4}{3}\lambda^3+x\lambda\bigr),
\end{equation}
and
\begin{equation}\label{Psi_at_0}
\|\Psi_{\mbox{\rm\tiny right}}(\lambda)\lambda^{-\alpha\sigma_3}\|\leq 
\mbox{\rm const},\qquad
\lambda\to0;
\end{equation}
\item
on the contour $\gamma$, the jump condition holds 
\begin{equation}\label{jump_cond}
\Psi_+(\lambda)=\Psi_-(\lambda)S(\lambda),
\end{equation}
where the piece-wise constant matrix $S(\lambda)$ is given by
equations:

-- on the rays $\gamma_k$,
\begin{equation}\label{p23}
S(\lambda)\bigr|_{\gamma_k}=S_k,\qquad
S_{2k-1}=I+s_{2k-1}\sigma_-,\qquad
S_{2k}=I+s_{2k}\sigma_+,
\end{equation}
with the constants $s_k$ satisfying the constraints
\begin{equation}\label{s_constraints}
s_{k+3}=-s_k,\qquad
s_1-s_2+s_3+s_1s_2s_3=-2\sin\pi\alpha;
\end{equation}

-- on the radiuses $\rho_{\pm}$, $S(\lambda)$ is specified by the equations
\begin{multline}\label{M_rels}
\lambda\in\rho_-\colon\
\Psi_{\mbox{\rm\tiny left}}(\rme^{2\pi i}\lambda)=
\Psi_{\mbox{\rm\tiny right}}(\lambda)M,
\\
\shoveleft{
\lambda\in\rho_+\colon\
\Psi_{\mbox{\rm\tiny right}}(\lambda)=
\Psi_{\mbox{\rm\tiny left}}(\lambda)\sigma_2M\sigma_2,
}\hfill
\end{multline}
where
\begin{multline}\label{M}
M=\Bigl(e^{i\pi(\alpha-\frac{1}{2})\sigma_3}
+J_{\pm}\sigma_{\pm}\Bigr)(-i\sigma_2),
\\
\shoveleft{\text{with}} 
\\
\shoveleft{ J_+=0\quad\hbox{ if }\quad 
\tfrac{1}{2}+\alpha\notin{\Bbb N},\quad
\hbox{i.e.}\quad
\alpha\notin\bigl\{\tfrac{1}{2},\tfrac{3}{2},\dots\bigr\}, } 
\\
\shoveleft{
J_-=0\quad\hbox{ if }\quad  \tfrac{1}{2}-\alpha\notin {\Bbb N},
\quad\hbox{i.e.}\quad
\alpha\notin\bigl\{-\tfrac{1}{2},-\tfrac{3}{2},\dots\bigr\};
}\hfill
\end{multline}

-- on the circle $C$, the piece-wise constant jump matrix $S(\lambda)$ is
defined by the equations
\begin{multline}\label{E_rels}
\Psi_0(\lambda)=\Psi_{\mbox{\rm\tiny right}}(\lambda)ES_0^{-1},\qquad
\Psi_1(\lambda)=\Psi_{\mbox{\rm\tiny right}}(\lambda)E,\qquad
\Psi_2(\lambda)=\Psi_{\mbox{\rm\tiny right}}(\lambda)ES_1,
\\
\shoveleft{
\Psi_3(\lambda)=
\Psi_{\mbox{\rm\tiny left}}(\lambda)\sigma_2E\sigma_2S_3^{-1},\qquad
\Psi_4(\lambda)=\Psi_{\mbox{\rm\tiny left}}(\lambda)\sigma_2E\sigma_2,
}
\\
\shoveleft{
\Psi_5(\lambda)=\Psi_{\mbox{\rm\tiny left}}(\lambda)\sigma_2E\sigma_2S_4.
}\hfill
\end{multline}
The connection matrix $E$, the Stokes matrices $S_k$ and the monodromy
matrix $M$ satisfy the cyclic relation,
\begin{equation}\label{E_conditions}
ES_1S_2S_3=\sigma_2M^{-1}E\sigma_2.
\end{equation}
\end{enumerate}

A solution of the RH problem (\ref{Psi_at_infty})--(\ref{E_conditions}), if
exists, is unique.

\begin{rem}\label{resonant}
For the $\lambda$-equation associated with PII (\ref{Lax_pair_p2_lambda}) as 
well as for the above RH problem, the values $\alpha=\frac{1}{2}+n$, 
$n\in{\Bbb Z}$, are called {\em resonant} since the corresponding 
$\Psi$-function may have a logarithmic singularity at the origin. It is 
a quite common fallacy, that the RH problem is {\em not} uniquely solvable 
for resonant values of such parameters. As a matter of facts, monodromy data 
form a locally smooth complex surface (\ref{s_constraints}) with the special 
points $\alpha=\frac{1}{2}+n$, $n\in{\Bbb Z}$, $s_1=-s_2=s_3=(-1)^{n+1}$. To
the latter, one has to attach a copy of ${\Bbb CP}^1$. A complex parameter 
describing the attached space ${\Bbb CP}^1$ can be interpreted as the ratio 
of a column entries of the connection matrix $E$ \cite{FN}. Thus neglecting 
the attached space ${\Bbb CP}^1$ may lead to the loss of uniqueness in the 
RH problem solution.
\end{rem}

The asymptotics of $\Psi(\lambda)$ as $\lambda\to\infty$ is given by
\begin{equation}\label{Y_expansion}
Y(\lambda):=\Psi(\lambda)e^{\theta\sigma_3}=
I+\lambda^{-1}\bigl(-i{\Eu H}\sigma_3+\frac{y}{2}\sigma_1\bigr)+
{\cal O}(\lambda^{-2}),
\end{equation}
where
\begin{equation}\label{H_def}
{\Eu H}=\tfrac{1}{2}y_x^2-\tfrac{1}{2}y^4-\tfrac{1}{2}xy^2+\alpha y
\end{equation}
is the Hamiltonian for the second Painlev\'e equation. Thus $y(x)$ can 
be extracted from the ``residue" of $Y(\lambda)$ at infinity,
\begin{equation}\label{y_from_Y}
y=2\lim_{\lambda\to\infty}\lambda Y_{12}(\lambda)=
2\lim_{\lambda\to\infty}\lambda Y_{21}(\lambda).
\end{equation}
Equation (\ref{y_from_Y}) specifies the Painlev\'e transcendent as the
function $y=f(x,\alpha,\{s_k\})$ of the independent variable $x$, of the
parameter $\alpha$ in the equation and of the Stokes multipliers $s_k$
(generically, the connection matrix $E$ can be expressed via $s_k$ using 
equation (\ref{E_conditions}) modulo arbitrary left diagonal (or triangular 
for half-integer $\alpha$) multiplier; at the special points of the 
monodromy surface (\ref{s_constraints}) $s_1=-s_2=s_3=(-1)^{n+1}$, 
$\alpha=\frac{1}{2}+n$, $n\in{\Bbb Z}$, the connection matrix $E$ 
contains a parameter $r\in{\Bbb C}P^1$ which specifies a relevant 
classical solution of $P_2$). Using the solution $y=f(x,\alpha,\{s_k\})$ 
and the symmetries of the Stokes multipliers described in \cite{kapaev:ell}, 
we obtain further solutions of $P_2$:
\begin{multline}\label{P_symmetries}
y=-f(x,-\alpha,\{-s_k\}),\qquad
y=\overline{f(\bar x,\bar\alpha,\{\overline{s_{4-k}}\})},
\\
\shoveleft{
y=\rme^{\rmi \frac{2\pi}{3}n}
f(\rme^{\rmi \frac{2\pi}{3}n}x,\alpha,\{s_{k+2n}\}),
}\hfill
\end{multline}
where the bar means the complex conjugation.

\section{Riemann-Hilbert problem for $1+s_0s_1=0$}

\begin{figure}[hbt]
\begin{center}
\epsfig{file=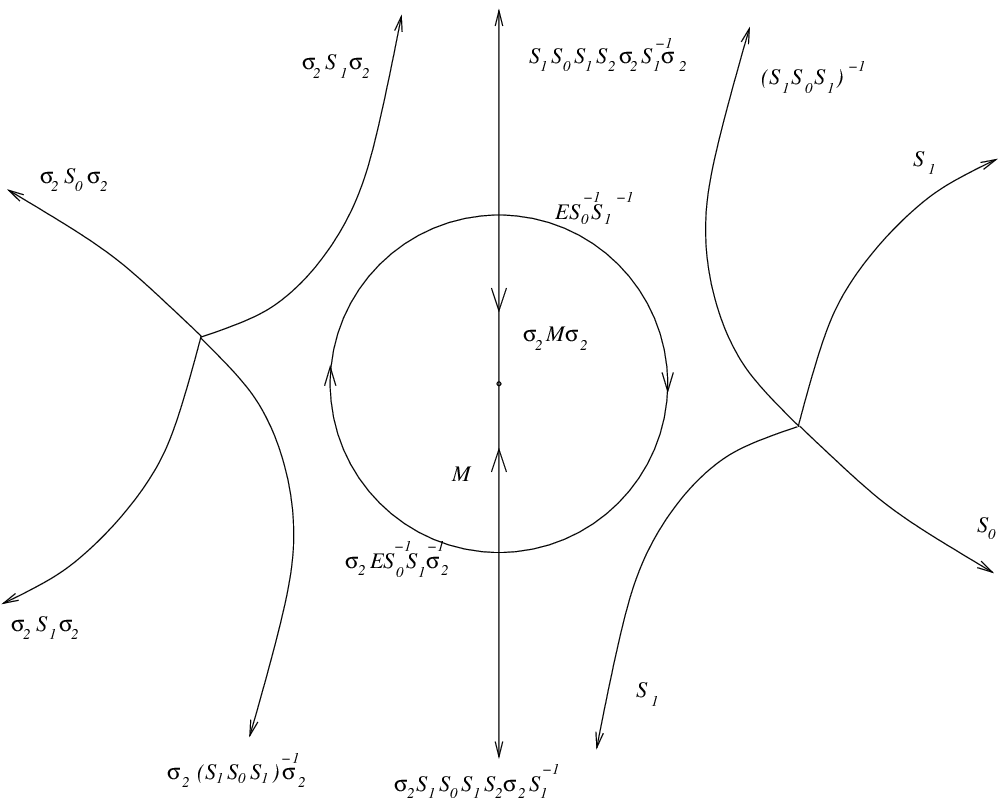}
\caption{A transformed RH problem graph.}
\label{fig2}
\end{center}
\end{figure}

First of all observe that our RH problem can be transformed to an equivalent 
RH problem with the jump contour consisting of three branches, see 
Figure~\ref{fig2} (the method of transformation of Riemann-Hilbert graphs is 
explained in detail in \cite{its_kapaev3, its_kapaev1}).

Let us assume now that
\begin{equation}\label{1+s0s1=0}
1+s_0s_1=0.
\end{equation} 
Constraints (\ref{s_constraints}) imply that
$s_1-s_0=-2\sin\pi\alpha$ as well, so that
\begin{equation}\label{s13}
s_1=e^{-i\pi\sigma(\alpha+\frac{1}{2})},\quad
s_0=-1/s_1=e^{i\pi\sigma(\alpha-\frac{1}{2})},\quad 
\sigma^2=1,
\end{equation}
while the parameter $s_2$ remains arbitrary. Using (\ref{1+s0s1=0}),
it is straightforward to check that the jump matrices in the transformed
jump graph in Figure~\ref{fig2} are as follows,
\begin{multline}\label{S1S0S1}
(S_1S_0S_1)^{-1}=
\begin{pmatrix}
0&1/s_1\\-s_1&0
\end{pmatrix},\quad
\sigma_2(S_1S_0S_1)^{-1}\sigma_2=
\begin{pmatrix}
0&s_1\\-1/s_1&0
\end{pmatrix},
\\
\shoveleft{
S_1S_0S_1S_2\sigma_2S_1^{-1}\sigma_2=
\begin{pmatrix}
0&-1/s_1\\s_1&s_1(s_1+s_2)
\end{pmatrix}=
\begin{pmatrix}
0&-1/s_1\\
s_1&0
\end{pmatrix}
\begin{pmatrix}
1&s_1+s_2\\
0&1
\end{pmatrix}
},
\\
\sigma_2S_1S_0S_1S_2\sigma_2S_1^{-1}=
\begin{pmatrix}
s_1(s_1+s_2)&-s_1\\1/s_1&0
\end{pmatrix}=
\begin{pmatrix}
0&-s_1\\
1/s_1&0
\end{pmatrix}
\begin{pmatrix}
1&0\\
-s_1-s_2&1
\end{pmatrix}.
\end{multline}

For the connection matrix $E$ we have:
\begin{subequations}\label{hatE}
\begin{multline}\label{hatEa}
\alpha-\tfrac{1}{2}\notin{\Bbb Z},\quad
J_+=J_-=0,\quad
M=e^{i\pi\alpha\sigma_3}i\sigma_1
\colon
\\
\begin{cases}
\sigma=-1,\quad
s_1=-e^{i\pi(\alpha-\frac{1}{2})}\colon\quad
ES_0^{-1}S_1^{-1}=p^{\sigma_3}
\begin{pmatrix}
1&0\\
\frac{s_1^2(s_1+s_2)}{1-s_1^2}&1
\end{pmatrix},
\\
\sigma=+1,\quad
s_1=-e^{-i\pi(\alpha-\frac{1}{2})}\colon\quad
ES_0^{-1}S_1^{-1}=p^{\sigma_3}
i\sigma_2
\begin{pmatrix}
1&0\\
\frac{s_1^2(s_1+s_2)}{1-s_1^2}&1
\end{pmatrix},
\end{cases}
\end{multline}
where $p$ is constant;
\begin{multline}\label{hatEb}
\alpha-\tfrac{1}{2}=n\in{\Bbb Z},\quad
s_1=-s_0=(-1)^{n+1},\quad
M\,i\sigma_2=(-1)^nI
+J_{\pm}\sigma_{\pm}
\colon\quad
\\
\begin{cases}
n\in{\Bbb Z}_{<0},\quad
J_+=0,\quad
J_-\neq0\colon\quad
ES_0^{-1}S_1^{-1}=p^{\sigma_3}
\begin{pmatrix}
1&0\\
q&1
\end{pmatrix}, 
\\ 
n\in{\Bbb Z}_{\geq0},\quad
J_+\neq0,\quad J_-=0\colon\quad
ES_0^{-1}S_1^{-1}=p^{\sigma_3}i\sigma_2
\begin{pmatrix}
1&0\\ q&1
\end{pmatrix},
\end{cases}
\end{multline}
\end{subequations}
where $p$ and $q$ are constants.

If $\alpha-\frac{1}{2}=n\in{\Bbb Z}$ and $s_1=-s_0=(-1)^{n+1}$ then the
condition $J_+=J_-=0$ is equivalent to the equation $s_2=(-1)^n$. On the one 
hand, these points have important geometrical meaning as being special points 
of the monodromy surface (\ref{s_constraints}). On the other hand, in this
case, equation (\ref{E_conditions}) does not provide any restriction to the 
connection matrix $E$ which becomes arbitrary. Taking into account that, for 
$\alpha-\frac{1}{2}=n\in{\Bbb Z}$, the matrix $E$ is determined up to an 
upper or lower in dependence on the sign of $n$ triangular left multiplier, 
it contains one essential parameter which parameterizes a family of the
(classical) Painlev\'e functions.

\section{Case $\alpha-\frac{1}{2}\not\in{\Bbb Z}$ and $1+s_0s_1=0$}
\label{case_a-1/2_not_Z}

\begin{figure}[hbt]
\begin{center}
\epsfig{file=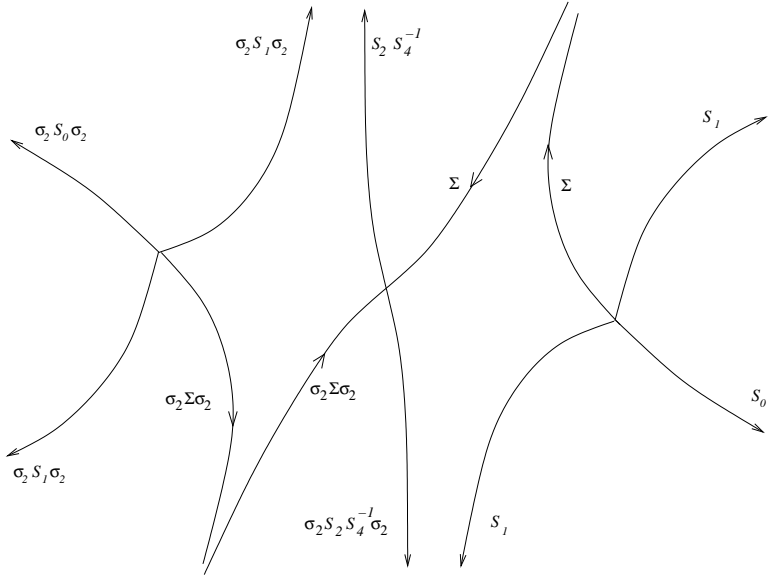}
\caption{An RH problem graph for non-special points of the
monodromy surface.}
\label{fig3}
\end{center}
\end{figure}

In the case of the non-special point of the monodromy surface, even for
a half-integer $\alpha$, the entries of the connection matrix $E$ do not
affect the Painlev\'e function, and thus the jump graph can be deformed 
to one depicted in Figure~\ref{fig3}.
Here,
\begin{equation}\label{Sigma_def}
\Sigma=
\begin{pmatrix}
0&1/s_1\\
-s_1&0
\end{pmatrix},\quad
S_2S_4^{-1}=
\begin{pmatrix}
1&s_1+s_2\\
0&1
\end{pmatrix},
\end{equation}
and other jump matrices are as above. For our convenience, we put the node
points of the jump graph to the points $\lambda_{1,2}=\pm\sqrt{-x/2}$, 
$\arg x\in\bigl[\frac{2\pi}{3},\pi\bigr]$. It is worth to note that the 
jump graph consists of the level lines $\Im g(\lambda)=const$, 
$\Re g(\lambda)=const$ emanating from the critical points 
$\lambda=\pm\sqrt{-x/2}$ and $\lambda=0$ for the function
\begin{equation}\label{g-8_def}
g(\lambda)=i\tfrac{4}{3}\bigl(\lambda^2+\tfrac{x}{2}\bigr)^{3/2}.
\end{equation}

The function $\Psi(\lambda)$ is normalized at infinity by the asymptotic
condition (\ref{Psi_at_infty}). Since we eliminate from the jump graph the
circle around the origin, the asymptotics for $\Psi(\lambda)$ at the origin 
(\ref{Psi_at_0}) has to be replaced by the condition
\begin{equation}\label{Psi_at_0_non_spec}
\bigl\|\Psi(\lambda)S_1S_0E^{-1}
\lambda^{-\alpha\sigma_3}\bigr\|\leq
\mbox{\rm const},\quad
\lambda\to+0.
\end{equation}

The RH problem depends on the free parameter $s_2$ due to a jump across 
the imaginary axis. Using a quadratic change $\lambda^2=\xi$ supplemented by
a gauge transformation of the $\Psi$-function, it is possible to obtain a
disjoint jump graph. Unfortunately, unlike the cases considered in 
\cite{its_kapaev1, kapaev2}, in this case, one meets a difficulty with the
normalization of the transformed RH problem which results in ambiguity in
the asymptotics of the Painlev\'e transcendent. Thus we prefer to deal with
the connected jump graph shown in Figure~\ref{fig3}.

\subsection{Reduced RH problem with $\bf 1+s_0s_1=s_1+s_2=0$.}

Consider the non-special RH problem for $P_2$ corresponding to 
$1+s_0s_1=s_1+s_2=0$. Then the non-specialty assumption implies that
$\alpha-\frac{1}{2}$ is not integer. The RH problem jump graph coincides 
with one depicted in Figure~\ref{fig3} except for the jump across the curve 
lines emanating from the origin and approaching the vertical direction 
since now
$$
S_2S_4^{-1}=\sigma_2S_2S_4^{-1}\sigma_2=I.
$$

The reduced RH problem is formulated as follows: find a piece-wise
holomorphic function $\hat\Psi(\lambda)$ such that
\begin{enumerate}
\item\quad
$\hat\Psi(\lambda)e^{\theta\sigma_3}\to I,\quad
\lambda\to\infty$;
\item\quad
$\bigl\|\hat\Psi_-(\lambda)S_1S_0E_0^{-1}
\lambda^{-\alpha\sigma_3}\bigr\|\leq
\mbox{\rm const},\quad
\lambda\to0$,\quad where $E_0$ is the connection matrix $E$ defined in 
(\ref{hatE}) corresponding to $s_1+s_2=0$;
\item\quad
$\hat\Psi(\lambda)$ is discontinuous across the contour shown in
Figure~\ref{fig3} (with the trivial jumps across the lines emanating from 
the origin and approaching the vertical directions).
\end{enumerate}

\begin{theorem}\label{McLeod1}
Let $\arg x\in[\frac{2\pi}{3},\pi]$, $\alpha-\frac{1}{2}\notin{\Bbb Z}$ and 
$1+s_0s_1=s_1+s_2=0$. Then, for large enough $|x|$, there exists a unique 
solution to the reduced RH problem. If additionally the parameter
$\sigma\in\{+1,-1\}$ is defined in such a way that
$s_1=e^{-i\pi\sigma(\alpha+\frac{1}{2})}$ then the corresponding Painlev\'e
function has the asymptotics 
\begin{equation}\label{y1_def}
y(x)=y_1(x,\alpha,\sigma)=
\sigma\sqrt{e^{-i\pi}\tfrac{x}{2}}+{\cal O}(x^{-2/5}),\quad
|x|\to\infty,\quad
\arg x\in\bigl[\tfrac{2\pi}{3},\pi\bigr].
\end{equation}
\end{theorem}

\proof 
Uniqueness. Let the reduced RH problem admit two solutions
$\hat\Psi_1(\lambda)$ and $\hat\Psi_2(\lambda)$. Since all the jump matrices
are unimodular, their determinants $\det\hat\Psi_j(\lambda)$ are continues 
across the jump contour, bounded at the origin and therefor are entire 
functions. Using the Liouville theorem and the normalization of 
$\hat\Psi(\lambda)$ we conclude that $\det\hat\Psi_j(\lambda)\equiv1$. 
Therefore there exists the ratio 
$\chi(\lambda)=\hat\Psi_1(\lambda)\hat\Psi_2^{-1}(\lambda)$ which is
continuous across the jump contour for $\hat\Psi_j(\lambda)$, remains bounded 
as $\lambda\to0$ and is normalized to the unit matrix as $\lambda\to\infty$.
Therefore, by the Liouville theorem, $\chi(\lambda)\equiv I$.

Existence.
Consider an RH problem on the curve line segment 
$[-\sqrt{-x/2},0)\cup(0,\sqrt{-x/2}]$, where
$\arg\sqrt{-x/2}\in\bigl[-\frac{\pi}{6},0\bigr]$, see Figure~\ref{fig3a} 
(because $\Psi$-function can be analytically continued to 
${\Bbb C}\backslash\{0\}$, the jump contours in the RH problem can be bent 
in any convenient way and even ``kiss" the infinity), with 
the quasi-permutation jump matrices:
\begin{figure}[hbt]\begin{center}
\epsfig{file=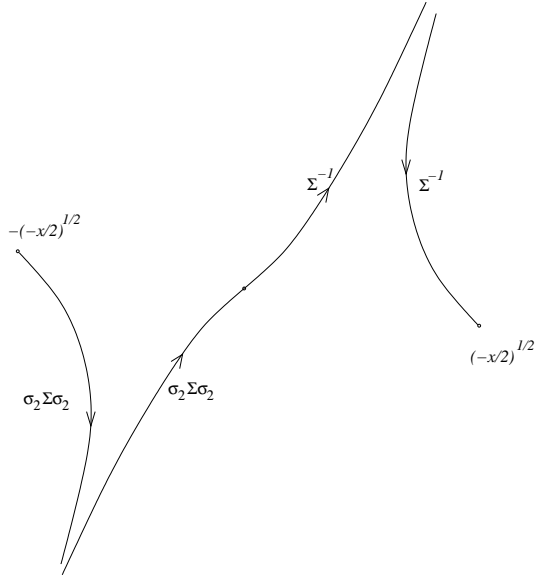}
\caption{Jump graph to the RH model problem 1.}
\label{fig3a}
\end{center}
\end{figure}

{\bf RH model problem 1.}
\begin{enumerate}
\item\quad
$\Phi^0(\lambda)e^{\theta\sigma_3}\to I,\quad
\lambda\to\infty$;
\item\quad
$\bigl\|\Phi^0_-(\lambda)S_1S_0E_0^{-1}
\lambda^{-\alpha\sigma_3}\bigr\|\leq
\mbox{\rm const},\quad
\lambda\to0$;
\item\quad
\begin{align*}
&\Phi^0_+(\lambda)=\Phi^0_-(\lambda)\sigma_2\Sigma\sigma_2,\quad
\lambda\in\bigl(-\sqrt{-x/2},0\bigr),
\\
&\Phi^0_+(\lambda)=\Phi^0_-(\lambda)\Sigma^{-1},\quad
\lambda\in\bigl(0,\sqrt{-x/2}\bigr),
\\
&\sigma_2\Sigma\sigma_2=
\begin{pmatrix}
0&s_1\\
-1/s_1&0
\end{pmatrix},\quad
\Sigma^{-1}=
\begin{pmatrix}
0&-1/s_1\\
s_1&0
\end{pmatrix}.
\end{align*}
\end{enumerate}

A solution to this RH problem is found explicitly in \cite{IFK} in some
different notations. Namely, on the complex $\lambda$-plane cut along
$[-\sqrt{-x/2},\sqrt{-x/2}]$, define the following scalar and matrix
functions $\beta,Y,\delta$:
\begin{equation}\label{beta_def}
\beta(\lambda)=\Bigl(
\frac{\lambda-\sqrt{-x/2}}{\lambda+\sqrt{-x/2}}
\Bigr)^{1/4},
\end{equation}
whose branch is fixed by the condition 
$$
\beta\to1\quad\text{as}\quad
\lambda\to\infty;
$$
\begin{equation}\label{Y_def}
Y(\lambda)=\frac{1}{2}
\begin{pmatrix}
\beta+\beta^{-1}&-\sigma(\beta-\beta^{-1})\\
-\sigma(\beta-\beta^{-1})&\beta+\beta^{-1}
\end{pmatrix}=
\tfrac{1}{\sqrt2}
\begin{pmatrix}
1&1\\1&-1
\end{pmatrix}
\beta^{-\sigma\sigma_3}
\tfrac{1}{\sqrt2}
\begin{pmatrix}
1&1\\1&-1
\end{pmatrix},
\end{equation}
\begin{equation}\label{delta_def}
\delta(\lambda)=\Bigl(
\frac{\sqrt{\lambda^2+\frac{x}{2}}-i\sqrt{-x/2}}
{\sqrt{\lambda^2+\frac{x}{2}}+i\sqrt{-x/2}}
\Bigr)^{\nu},\quad
\nu=-\frac{1}{2\pi i}\ln(i\sigma s_1)=\frac{\sigma\alpha}{2},
\end{equation}
where the branch of the square root is fixed by its asymptotics
$\sqrt{\lambda^2+\frac{x}{2}}=\lambda+{\cal O}(\lambda^{-1})$ as 
$\lambda\to+\infty$, and $z^{\nu}$ is defined on the plane cut along 
the negative part of the real axis and its main branch is chosen.

As easy to see,
\begin{equation}\label{delta_symm}
\delta(-\lambda)\delta(\lambda)=1.
\end{equation}

Furthermore, the functions introduced above enjoy the following jump properties:
\begin{align}\label{Y_jump}
&Y_+(\lambda)=Y_-(\lambda)(-i\sigma)\sigma_1,\quad
\lambda\in\bigl(-\sqrt{-x/2},\sqrt{-x/2}\bigr),
\\\label{delta_jump}
&\delta_+(\lambda)\delta_-(\lambda)=e^{2\pi i\nu},\quad
\lambda\in\bigl(-\sqrt{-x/2},0\bigr),
\\
&\delta_+(\lambda)\delta_-(\lambda)=e^{-2\pi i\nu},\quad
\lambda\in\bigl(0,\sqrt{-x/2}\bigr),
\notag
\\\label{g_jump}
&g_+(\lambda)=-g_-(\lambda),\quad
\lambda\in\bigl(-\sqrt{-x/2},\sqrt{-x/2}\bigr).
\end{align}
Therefore the function
\begin{equation}\label{Phi0_def}
\Phi^0=Y\delta^{\sigma_3}e^{-g\sigma_3}
\end{equation}
satisfies the jump conditions of model RH problem~1.

Because of the asymptotics as $\lambda\to\infty$
\begin{align}\label{beta_Y_delta_g_as8}
&\beta(\lambda)=1-\frac{1}{2\lambda}\sqrt{-x/2}+{\cal O}(\lambda^{-2}),
\notag
\\
&Y(\lambda)=I+\frac{1}{2\lambda}\sigma\sqrt{-\tfrac{x}{2}}\,\sigma_1
+{\cal O}(\lambda^{-2}),
\notag
\\
&\delta(\lambda)=1+{\cal O}(\lambda^{-2}),
\notag
\\
&g(\lambda)=i\bigl(
\tfrac{4}{3}\lambda^3+x\lambda\bigr)
+i\tfrac{x^2}{8\lambda}
+{\cal O}(\lambda^{-3}),
\end{align}
we find the asymptotics as $\lambda\to\infty$ of $\Phi^0(\lambda)$,
\begin{equation}\label{Phi0_as8}
\Phi^0(\lambda)=\Bigl(
I+\tfrac{1}{2\lambda}
\bigl(-\tfrac{ix^2}{4}\sigma_3+\sigma\sqrt{-\tfrac{x}{2}}\,\sigma_1\bigr)
+{\cal O}(\lambda^{-2})\Bigr)
e^{-\theta\sigma_3}.
\end{equation}
Using asymptotics as $\lambda\to0$,
\begin{align}\label{beta_Y_delta_g_as0}
&\beta_-(\lambda)=e^{-i\pi/4}+{\cal O}(\lambda),
\notag
\\
&Y_-(\lambda)=\frac{1}{\sqrt2}
\Bigl(I+i\sigma\sigma_1+{\cal O}(\lambda)\Bigr),
\notag
\\
&\delta_-(\lambda)=e^{-i\pi\nu}(-2x)^{\nu}\lambda^{-2\nu}
\bigl(1+{\cal O}(\lambda^2)\bigr),
\notag
\\
&g_-(\lambda)=-\tfrac{\sqrt2}{3}(-x)^{3/2}+\sqrt2(-x)^{1/2}\lambda^2
+{\cal O}(\lambda^4x^{-1/2}),
\end{align}
it is easy to see that
\begin{equation}\label{Phi0_as0}
\Phi^0_-(\lambda)=C\bigl(I+{\cal O}(\lambda)\bigr)
\lambda^{-\sigma\alpha\sigma_3}
\end{equation}
with some constant matrix $C$, therefore, using (\ref{hatEa}),
$\bigl\|\Phi^0_-(\lambda)S_1S_0E_0^{-1}
\lambda^{-\alpha\sigma_3}\bigr\|\leq\mbox{\rm const}$. Thus the function
$\Phi^0(\lambda)$ (\ref{Phi0_def}) solves model RH problem~1.

\begin{rem}\label{quasi_perm}
A similar quasi-permutation RH problem in an elliptic case is solved in
\cite{its_kapaev3}. More general quasi-permutation RH problem is solved in 
\cite{korotkin}. 
\end{rem}

For the subsequent discussion it is also worth to note that
\begin{equation}\label{Phi0_sym}
\sigma_2\Phi^0(-\lambda)\sigma_2=\Phi^0(\lambda).
\end{equation}
We also point out that $\Phi^0(\lambda)$ is singular at 
$\lambda=\pm\sqrt{-x/2}$. To ``smoothen" this singularity, 
let us consider the following

\begin{figure}[hbt]\begin{center}
\epsfig{file=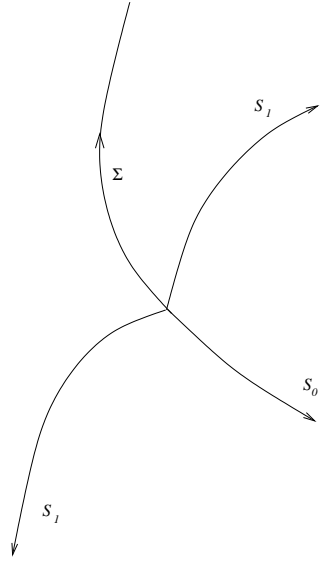}
\caption{RH model problem near $\lambda=\sqrt{-x/2}$.}
\label{fig3b}
\end{center}
\end{figure}

{\bf RH model problem 2.}
Find a piece-wise holomorphic function $\Phi(\lambda)$ with the jumps
indicated in Figure~\ref{fig3b}.
Since we do not normalize $\Phi(\lambda)$, it is determined up to a left
multiplication in an entire matrix function. If we will use this solution in 
a domain different from ${\Bbb C}$, then $\Phi(\lambda)$ is determined up to 
a left multiplication in a matrix function holomorphic in this domain.

It is possible to construct a solution $\Phi(\lambda)$ of this model problem
using the classical Airy functions. The Airy function $\Ai(z)$ can be 
defined using the Taylor expansion \cite{BE, olver},
\begin{equation}\label{Ai}
\Ai(z)=\frac{1}{3^{2/3}\Gamma(\frac{2}{3})} 
\sum_{k=0}^{\infty}
\frac{3^k\Gamma(k+\frac{1}{3})z^{3k}}{\Gamma(\frac{1}{3})(3k)!}
-\frac{1}{3^{1/3}\Gamma(\frac{1}{3})} 
\sum_{k=0}^{\infty}
\frac{3^k\Gamma(k+\frac{2}{3})z^{3k+1}}{\Gamma(\frac{2}{3})(3k+1)!}.
\end{equation}
Asymptotics at infinity of this function and its derivative are as
follows,
\begin{multline}\label{Ai_as}
\Ai(z)=\tfrac{z^{-1/4}}{2\sqrt\pi}e^{-\frac{2}{3}z^{3/2}}
\Bigl\{\sum_{n=0}^N(-1)^n3^{-2n}
\frac{\Gamma(3n+\frac{1}{2})}{\Gamma(\frac{1}{2})(2n)!}z^{-\frac{3n}{2}}
+{\cal O}\bigl(z^{-\frac{3}{2}(N+1)}\bigr)\Bigr\}, 
\\
\Ai'(z)=-\tfrac{z^{1/4}}{2\sqrt\pi}e^{-\frac{2}{3}z^{3/2}} 
\Bigl\{\sum_{n=0}^N(-1)^{n+1}3^{-2n}\bigl(3n+\tfrac{1}{2}\bigr)
\frac{\Gamma(3n-\frac{1}{2})}{\Gamma(\frac{1}{2})(2n)!}z^{-\frac{3n}{2}}
+{\cal O}\bigl(z^{-\frac{3}{2}(N+1)}\bigr) 
\Bigr\}, 
\\ 
\text{as}\quad
z\to\infty,\quad 
\arg z\in(-\pi,\pi).
\end{multline}

Introduce the matrix function $Z_0(\lambda)$,
\begin{equation}\label{Z0_def}
Z_0(\lambda)=\sqrt{2\pi}\,e^{-i\pi/4}
\begin{pmatrix}
v_2(z)&v_1(z)\\
\frac{d}{dz}v_2(z)&\frac{d}{dz}v_1(z)
\end{pmatrix}e^{-i\frac{\pi}{4}\sigma_3},
\end{equation}
where
\begin{equation}\label{Airy_notations}
v_2(z)=e^{i2\pi/3}Ai(e^{i2\pi/3}z),\quad
v_1(z)=Ai(z).
\end{equation}
As $|z|\to\infty$, $\arg z\in\bigl(-\pi,\frac{\pi}{3}\bigr)$, this function
has the asymptotics
\begin{equation}\label{Z_0_as}
Z_0(z)=z^{-\sigma_3/4}\frac{1}{\sqrt2}
\begin{pmatrix}
1&1\\
1&-1
\end{pmatrix}
\Bigl(I+{\cal O}(z^{-3/2})\Bigr)
e^{\frac{2}{3}z^{3/2}\sigma_3}. 
\end{equation}
Also, introduce the matrix functions
\begin{multline}\label{Airies}
Z_1(z):=Z_0(z)G_0,\quad
Z_2(z):=Z_1(z)G_1,\quad
Z_3(z):=Z_2(z)G_2,
\\
\shoveleft{
G_0=G_2=
\begin{pmatrix}
1&0\\
-i&1
\end{pmatrix},\quad
G_1=
\begin{pmatrix}
1&-i
\\
0&1
\end{pmatrix}.
}\hfill
\end{multline}
By the properties of the Airy functions \cite{BE, olver}, $Z_j(z)$ have 
the asymptotics (\ref{Z_0_as}) as $z\to\infty$ and 
$\arg z\in\bigl(-\pi+\frac{2\pi}{3}j,\frac{\pi}{3}+\frac{2\pi}{3}j\bigr)$,
$j=0,1,2,3$.

\begin{figure}[hbt]\begin{center}
\epsfig{file=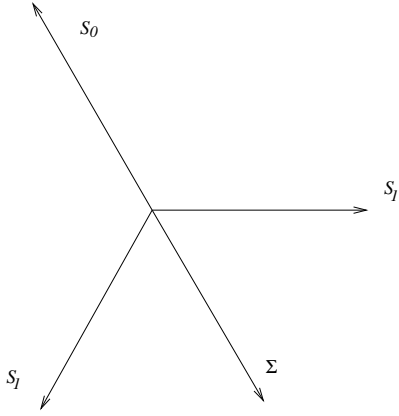}
\caption{A model RH problem solvable by the Airy functions.}
\label{fig5}
\end{center}
\end{figure}

Define a piece-wise holomorphic function $Z^{RH}(z)$,
\begin{equation}\label{RH_Airy}
Z^{RH}(z)=\begin{cases}
Z_0(z)(is_1)^{\sigma_3/2},\quad
\arg z\in\bigl(-\frac{\pi}{3},0),\\
Z_j(z)(is_1)^{\sigma_3/2},\quad
\arg z\in\bigl(-\frac{2\pi}{3}+\frac{2\pi}{3}j,\frac{2\pi}{3}j\bigr),
\quad
j=1,2,\\
Z_3(z)(is_1)^{\sigma_3/2},\quad
\arg z\in\bigl(\frac{4\pi}{3},\frac{5\pi}{3}).
\end{cases}
\end{equation}
By construction, $Z^{RH}(z)$ has the uniform asymptotics 
\begin{equation}\label{Z_RH_as}
Z^{RH}(z)=z^{-\sigma_3/4}\frac{1}{\sqrt2}
\begin{pmatrix}
1&1\\
1&-1
\end{pmatrix}
\Bigl(I+{\cal O}(z^{-3/2})\Bigr)
e^{\frac{2}{3}z^{3/2}\sigma_3}
(is_1)^{\sigma_3/2},\quad
z\to\infty,
\end{equation}
and jump properties indicated in Figure~\ref{fig5}.

Using the change of the independent variable
\begin{equation}\label{z(lambda)}
z=e^{i\pi}2^{2/3}\bigl(\lambda^2+\frac{x}{2}\bigr),
\end{equation}
and the notation 
\begin{equation}\label{zeta_def}
\zeta:=\lambda-\sqrt{-x/2},
\end{equation}
we compute the ``ratio" of $\Phi^0$ and $Z^{RH}$
in the annulus 
$c_1|x|^{-\frac{1}{2}+\epsilon}\leq|\zeta|\leq c_2|x|^{-\frac{1}{2}+\epsilon}$,
where $0<c_1<c_2$ and $\frac{1}{4}<\epsilon<\frac{1}{2}$ are some constants,
\begin{multline}\label{Phi0_Z_ratio}
\Phi^0(\lambda)\bigl(Z^{RH}(z)\bigr)^{-1}=
\Bigl(I+{\cal O}\bigl(x^{-1/4}\zeta^{1/2}\bigr)
+{\cal O}\bigl(x^{-1/2}\zeta^{-2}\bigr)\Bigr)
\times
\\
\times\tfrac{1}{\sqrt2}
\begin{pmatrix}
1&1\\1&-1
\end{pmatrix}
\Bigl(\frac{1-\sigma}{2}i\sigma_1
+\frac{1+\sigma}{2}I\Bigr)
e^{i\frac{\pi}{4}\sigma_3}
2^{\frac{2}{3}\sigma_3}
(-x/2)^{\sigma_3/4}.
\end{multline}

Define the matrix function $\Phi_r(\lambda)$,
\begin{equation}\label{Phir_def}
\Phi_r(\lambda)=
\tfrac{1}{\sqrt2}
\begin{pmatrix}
1&1\\1&-1
\end{pmatrix}
\Bigl(\frac{1-\sigma}{2}i\sigma_1
+\frac{1+\sigma}{2}I\Bigr)
e^{i\frac{\pi}{4}\sigma_3}
2^{\frac{2}{3}\sigma_3}
(-x/2)^{\sigma_3/4}
Z^{RH}\bigl(z(\lambda)\bigr).
\end{equation}

\begin{figure}[hbt]\begin{center}
\epsfig{file=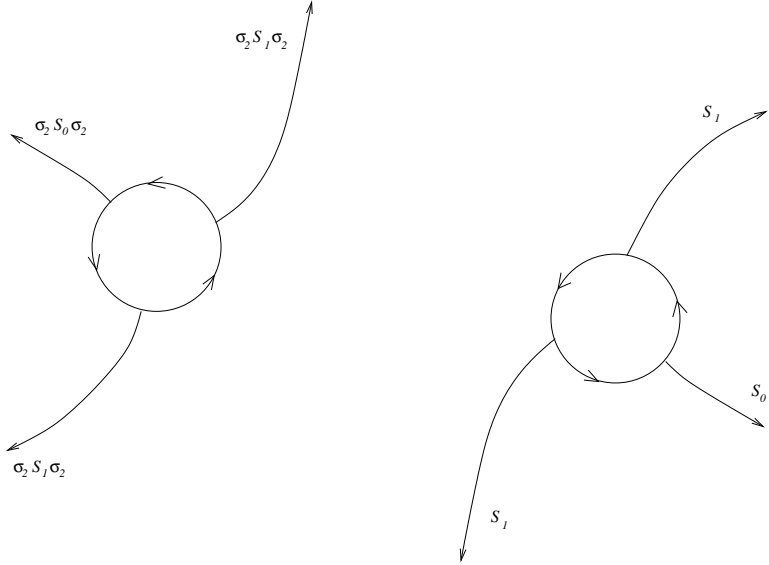}
\caption{RH problem jump graph for the correction function $\chi(\lambda)$.}
\label{fig6}
\end{center}
\end{figure}

Let us introduce the piece-wise holomorphic function $\Phi(\lambda)$,
\begin{equation}\label{Phi_def}
\Phi(\lambda)=
\begin{cases}
\Phi_r(\lambda),\quad
\bigl|\lambda-\sqrt{-x/2}\bigr|<R,\\
\sigma_2\Phi_r(e^{-i\pi}\lambda)\sigma_2,\quad
\bigl|\lambda+\sqrt{-x/2}\bigr|<R,\\
\Phi^0(\lambda),\quad
\bigl|\lambda\pm\sqrt{-x/2}\bigr|>R,
\end{cases}
\end{equation}
where $R=c|x|^{-\frac{1}{2}+\epsilon}$ for a constant $c>0$ and
$\frac{1}{4}<\epsilon<\frac{1}{2}$. We look for the exact solution of
the reduced RH problem in the form of the product
\begin{equation}\label{chi_red_def}
\hat\Psi(\lambda)=\chi(\lambda)\Phi(\lambda).
\end{equation}
The correction function $\chi(\lambda)$ solves the RH problem whose jump
graph is shown in Figure~\ref{fig6}.
The jump matrix across the circle centered at $\lambda=\sqrt{-x/2}$ is given 
by the ratio $\Phi^0(\lambda)\Phi_r^{-1}(\lambda)$ which, due to
(\ref{Phi0_Z_ratio}), (\ref{Phir_def}) and choosing $\epsilon=2/5$, 
satisfies the estimate
\begin{equation}\label{jump_r_circle}
\bigl\|\Phi^0(\lambda)\Phi_r^{-1}(\lambda)-I\bigr\|\leq
c|x|^{-3/10}.
\end{equation}
The jump matrices across the exterior parts of infinite contours approach
the unit matrices exponentially fast,
\begin{equation}\label{jump_tails}
\bigl\|\Phi^0(\lambda)S_j\bigl(\Phi^0(\lambda)\bigr)^{-1}-I\bigr\|\leq
C|x|^{3/20}e^{-\frac{2^{15/4}}{3}|x|^{3/4}|\zeta|^{3/2}}.
\end{equation}

Now, the solvability of the reduced RH problem is straightforward. Indeed,
consider the system of singular integral equations $\chi_-=I+K\chi_-$ 
equivalent to the RH problem for $\chi(\lambda)$,
\begin{equation}\label{chi_singular_integral}
\chi_-(\lambda)=I+\frac{1}{2\pi i}\int_{\gamma}\chi_-(\xi)
\bigl(G(\xi)-I\bigr)\frac{d\xi}{\xi-\lambda_-},
\end{equation}
where $\gamma$ is the contour shown in Figure~\ref{fig6} and
$G(\lambda)=\Phi_-(\lambda)S_j\Phi_+^{-1}(\lambda)$ is the relevant jump
matrix. Using the estimates (\ref{jump_r_circle}), (\ref{jump_tails}) and 
the boundedness of the Cauchy operator in $L^2(\gamma)$, the singular
integral operator $K$, which is a superposition of the operator of the right 
multiplication in $G(\lambda)-I$ and of the Cauchy operator $C_-$, satisfies
the estimate, $\bigl\|K\bigr\|_{L^2(\gamma)}\leq c|x|^{-2/5}$, where
the precise value of the positive constant $c$ is not important for us. Thus 
$K$ is a contracting operator in $L^2(\gamma)$ for large enough $|x|$. 
Since $KI$ is a square integrable function on $\gamma$, and observing that 
$\rho=\chi_--I$ satisfies the singular integral equation $\rho=KI+K\rho$, 
we find $\rho\in L^2(\gamma)$ and therefore $\chi_-=I+\rho$ by iterations.

The solution found by iterations implies the asymptotics of $\chi(\lambda)$
as $\lambda\to\infty$,
\begin{equation}\label{chi_as}
\chi(\lambda)=I+\frac{1}{\lambda}{\cal O}(x^{-2/5})
+{\cal O}(\lambda^{-2}).
\end{equation}
Using (\ref{chi_as}) and (\ref{Phi0_as8}) in (\ref{chi_red_def}) and taking
into account the expansion (\ref{Y_expansion}), we see that the Painlev\'e
function corresponding to the reduced RH problem has the asymptotics as
$|x|\to\infty$, $\arg x\in\bigl[\frac{2\pi}{3},\pi\bigr]$,
\begin{equation}\label{y_as}
y(x)=\sigma\sqrt{e^{-i\pi}\frac{x}{2}}+{\cal O}(x^{-2/5}).
\end{equation}
This completes the proof.
\endproof

\subsection{RH problem with $\bf 1+s_0s_1=0$ and $\bf s_1+s_2\neq0$.}

Let us go to the case of $1+s_0s_1=0$ and arbitrary $s_1+s_2\neq0$, see
Figure~\ref{fig3}. We look for the solution $\Psi(\lambda)$ in
the form of the product
\begin{equation}\label{hat_chi_def}
\Psi(\lambda)=\hat\chi(\lambda)\hat\Psi(\lambda),
\end{equation}
where $\hat\Psi(\lambda)$ is the solution of the reduced RH problem, i.e.\
with $s_1+s_2=0$. 

\begin{figure}[hbt]\begin{center}
\epsfig{file=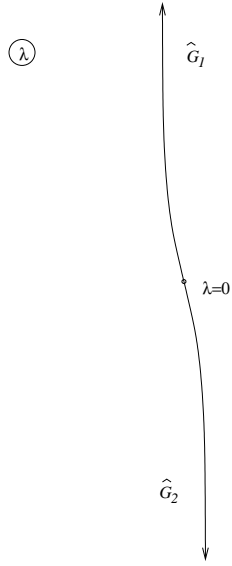}
\caption{An RH problem graph for the correction function
$\hat\chi(\lambda)$.}
\label{fig7}
\end{center}
\end{figure}

The correction function $\hat\chi(\lambda)$ satisfies the RH problem on two
level lines $\Im g(\lambda)=const$ emanating from the origin and approaching 
the vertical direction, see Figure~\ref{fig7},
\begin{enumerate}
\item\quad
$\hat\chi(\lambda)\to I,\quad 
\lambda\to\infty$;
\item\quad
$\bigl\|\hat\chi(\lambda)\hat\Psi(\lambda)S_1S_0E^{-1}
\lambda^{-\alpha\sigma_3}\bigr\|\leq\mbox{\rm const},\quad
\lambda\to+0$;
\item\quad
$\hat\chi_+(\lambda)=\hat\chi_-(\lambda)\hat G(\lambda)$,

\noindent\quad
$\hat G(\lambda):=\hat G_1(\lambda)=
\hat\Psi_-(\lambda)S_2S_4^{-1}\hat\Psi_-^{-1}(\lambda),\quad
\lambda\in(0,+i\infty)$,
\par
\noindent
\quad
$\hat G(\lambda):=\hat G_2(\lambda)=
\hat\Psi_-(\lambda)\sigma_2S_2S_4^{-1}\sigma_2\hat\Psi_-^{-1}(\lambda),\quad
\lambda\in(0,-i\infty)$,
\par
\noindent
\quad
$\sigma_2\hat G_1(-\lambda)\sigma_2=\hat G_2(\lambda)$.
\end{enumerate}

\begin{theorem}\label{McLeod2}
If $\alpha-\frac{1}{2}\notin{\Bbb Z}$, $1+s_0s_1=0$, 
$\arg x\in[\frac{2\pi}{3},\pi]$ and $|x|$ is large enough, 
then the RH problem (\ref{Psi_at_infty})--(\ref{E_conditions}) is uniquely 
solvable. The asymptotics of the corresponding Painlev\'e function as 
$|x|\to\infty$ in the indicated sector is given by
\begin{multline}\label{mcleod-_as}
y=y_1(x,\alpha,\sigma)-
\\
-\frac{s_1+s_2}{\pi}
2^{-\frac{5}{2}\sigma\alpha-\frac{7}{4}}
\Gamma(\tfrac{1}{2}+\sigma\alpha)
(e^{-i\pi}x)^{-\frac{3}{2}\sigma\alpha-\frac{1}{4}}
e^{-\frac{2\sqrt2}{3}(e^{-i\pi}x)^{3/2}}
\bigl(1+{\cal O}(x^{-1/4})\bigr),
\end{multline}
where $y_1(x,\alpha,\sigma)\sim\sigma\sqrt{e^{-i\pi}x/2}$ is the solution 
of the Painlev\'e equation corresponding to $1+s_0s_1=s_1+s_2=0$, while the
parameter $\sigma\in\{+1,-1\}$ is defined by the use of the equation 
$s_1=e^{-i\pi\sigma(\alpha+\frac{1}{2})}$.
\end{theorem}

\proof 
It is enough to prove the solvability of the RH problem for
$\hat\chi(\lambda)$ above. In a neighborhood of the imaginary axis,
$\hat\Psi(\lambda)=\chi(\lambda)\Phi^0(\lambda)$, see (\ref{chi_red_def})
and (\ref{Phi_def}). Hence, using (\ref{Phi0_def}), we have the following 
expression for the jump matrix across $(0,+i\infty)$,
\begin{subequations}\label{hat_G12}
\begin{align}\label{hat_G1}
&\hat G_1(\lambda)=I+(s_1+s_2)\delta^2e^{-2g}\chi Y\sigma_+Y^{-1}\chi^{-1},
\\\label{hat_G2}
&\hat G_2(\lambda)=I-(s_1+s_2)\delta^{-2}e^{2g}\chi Y\sigma_-Y^{-1}\chi^{-1}.
\end{align}
\end{subequations}
Because the jump contour coincides with the level line 
$\Im g(\lambda)=const$, the jump matrix $\hat G_1(\lambda)$ exponentially 
approaches the unit matrix as $\lambda\to+i\infty$. Similarly,
$\hat G_2(\lambda)-I$ decreases exponentially as $\lambda\to-i\infty$. For
the exponential $e^{g(\lambda)}$, due to (\ref{beta_Y_delta_g_as0}), 
the origin is the saddle point. 

The jump matrices have algebraic singularity at $\lambda=0$,
\begin{subequations}\label{hat_G12_at0}
\begin{align}\label{hat_G1_at0}
&\hat G_1(\lambda)=I+(s_1+s_2)e^{-2\pi i\nu}
(-2x)^{-2\nu}\lambda^{4\nu}
e^{-\frac{2\sqrt2}{3}(-x)^{3/2}+2\sqrt2(-x)^{1/2}\lambda^2}\times
\notag
\\
&\times
\tfrac{1}{2}(i\sigma\sigma_3+\sigma_1)
\bigl(1+{\cal O}(\lambda)+{\cal O}(x^{-9/10})\bigr),
\\\label{hat_G2_at0}
&\hat G_2(\lambda)=I-(s_1+s_2)
e^{2i\pi\nu}(-2x)^{-2\nu}\lambda^{4\nu}
e^{-\tfrac{2\sqrt2}{3}(-x)^{3/2}+2\sqrt2(-x)^{1/2}\lambda^2)}\times
\notag
\\
&\times
\tfrac{1}{2}(i\sigma\sigma_3+\sigma_1)
\bigl(1+{\cal O}(\lambda)+{\cal O}(x^{-9/10})\bigr).
\end{align}
\end{subequations}

Consider the model RH problem with the jump matrices $\tilde G_j(\lambda)$,
$j=1,2$, of the form (\ref{hat_G12}) but with the matrix
$\chi(\lambda)Y(\lambda)$ replaced by its asymptotics at $\lambda=0$, i.e.
\begin{subequations}\label{tilde_G12}
\begin{align}\label{tilde_G1}
&\tilde G_1(\lambda)=I+(s_1+s_2)\delta^2e^{-2g}
\tfrac{1}{2}(i\sigma\sigma_3+\sigma_1),
\\\label{tilde_G2}
&\tilde G_2(\lambda)=I-(s_1+s_2)\delta^{-2}e^{2g}
\tfrac{1}{2}(i\sigma\sigma_3+\sigma_1).
\end{align}
\end{subequations}

Assume for a moment that
\begin{equation}\label{nu>=0}
\Re\nu\geq0,\quad
\text{i.e.}\quad
\Re(\sigma\alpha)\geq0.
\end{equation}
Since $\tilde G_j(\lambda)-I$ are constant nilpotent matrices multiplied in
scalar functions, solution of the jump problem is given by the Cauchy integral,
\begin{multline}\label{tilde_chi_sol}
\tilde\chi(\lambda)=I+\frac{s_1+s_2}{2\pi i}
\Bigl\{
\int_0^{+i\infty}
\delta^2e^{-2g}
\frac{d\zeta}{\zeta-\lambda}
+\int_{-i\infty}^0
\delta^{-2}e^{2g}
\frac{d\zeta}{\zeta-\lambda}
\Bigr\}
\tfrac{1}{2}(i\sigma\sigma_3+\sigma_1)=
\\
=I+\frac{s_1+s_2}{2\pi i}
\int_0^{+i\infty}
\delta^2e^{-2g}
\frac{\lambda\,d\zeta}
{\zeta^2-\lambda^2}
(i\sigma\sigma_3+\sigma_1).
\end{multline}
The condition (\ref{nu>=0}) ensures the convergence of the above integrals.
Obviously, $\tilde\chi(\lambda)\to I$ as $\lambda\to\infty$. As to
asymptotics at the origin, the same condition (\ref{nu>=0}) ensures that
$\bigl\|\tilde\chi(\lambda)\bigr\|\leq\mbox{\rm const}$, and
$\bigl\|\tilde\chi(\lambda)\hat\Psi(\lambda)S_1S_0E^{-1}
\lambda^{-\alpha\sigma_3}\bigr\|\leq\mbox{\rm const}$.

We will look for $\hat\chi(\lambda)$ in the form of the product
\begin{equation}\label{X_def}
\hat\chi(\lambda)=X(\lambda)\tilde\chi(\lambda).
\end{equation}
For the correction function $X(\lambda)$, we have the RH problem with the
jump contour shown in Figure~\ref{fig7}, but with different jump matrices:
\begin{enumerate}
\item\quad
$X(\lambda)\to I,\quad 
\lambda\to\infty$;
\item\quad
$\bigl\|X(\lambda)\bigr\|\leq\mbox{\rm const},\quad
\lambda\to+0$;
\item\quad
$X_+(\lambda)=X_-(\lambda)H(\lambda)$,\quad
$H(\lambda)=\tilde\chi_-(\lambda)\hat G(\lambda)\tilde
G^{-1}(\lambda)\tilde\chi_-^{-1}(\lambda)$.
\end{enumerate}

In the equivalent singular integral equation,
\begin{equation}\label{X_sing_eq}
X_-(\lambda)=I+\frac{1}{2\pi i}\int_{\gamma}
X_-(\zeta)\bigl(H(\zeta)-I\bigr)\frac{d\zeta}{\zeta-\lambda_-},
\end{equation}
where $\gamma$ is the jump graph in Figure~\ref{fig7} and $\lambda_-$ is the
right limit of $\lambda$ on the jump contour, or, in the symbolic form,
$X_-=I+KX_-$, the operator $K$ is the superposition of the right
multiplication in $H-I$ and of the Cauchy operator $C_-$. Using the
boundedness of $C_-$ in $L^2(\gamma)$, we estimate the norm
\begin{equation}\label{K_H-I_norm}
\bigl\|K\bigr\|_{L^2(\gamma)}\leq
c\bigl\|H-I\bigr\|_{L^2(\gamma)}\leq
c'|x|^{-(5\nu+1)/2}e^{-\frac{8}{3}|x|^{3/2}\cos\frac{3}{2}(\arg x-\pi)}\leq
c''|x|^{-1/2},
\end{equation}
where $c$, $c'$ and $c''$ are positive constants whose precise value is not
important for us. Taking into account the assumption $\Re\nu\geq0$
(\ref{nu>=0}), the singular integral operator $K$ is contracting for large
enough $|x|$, $\arg x\in\bigl[\frac{2\pi}{3},\pi\bigr]$, and therefore
equation (\ref{X_sing_eq}) is solvable by iterations. 

To find the asymptotics of the relevant Painlev\'e function, it is enough to
find asymptotics of $\tilde\chi(\lambda)$ (\ref{tilde_chi_sol}) as
$\lambda\to\infty$,
\begin{multline}\label{tilde_chi_as}
\tilde\chi(\lambda)=I-\frac{s_1+s_2}{2\pi i\lambda}
\int_0^{+i\infty}\delta^2e^{-2g}\,d\zeta
\bigl(1+{\cal O}(\lambda^{-2})\bigr)
(i\sigma\sigma_3+\sigma_1)=
\\
=I+\lambda^{-1}h(i\sigma\sigma_3+\sigma_1)+{\cal O}(\lambda^{-3}),
\\
h=-\frac{s_1+s_2}{2\pi}
2^{-5\nu-\frac{7}{4}}
\Gamma(2\nu+\tfrac{1}{2})
(-x)^{-3\nu-\frac{1}{4}}
e^{-\frac{2\sqrt2}{3}(-x)^{3/2}}
\bigl(1+{\cal O}(x^{-1/2})\bigr).
\end{multline}
Thus the asymptotics of $\Psi(\lambda)$ as $\lambda\to\infty$
(\ref{Y_expansion}) is given by
\begin{multline}\label{Psi_as_s1+s2ne0}
\Psi(\lambda)e^{\theta\sigma_3}=X\tilde\chi\hat\Psi e^{\theta\sigma_3}=
I+\lambda^{-1}\bigl(-i{\Eu H}\sigma_3+\frac{y}{2}\sigma_1\bigr)+
{\cal O}(\lambda^{-2})=
\\
=
I+\lambda^{-1}\Bigl(
-i({\Eu H}_2-\sigma\hat h_3)\sigma_3
+\frac{y_1+2\hat h_1}{2}\sigma_1
\Bigr)+
{\cal O}(\lambda^{-2}),
\end{multline}
where $\hat h_j=h\bigl(1+{\cal O}(x^{-2\nu-1/4})\bigr)$, $j=3,1$, involves
the contribution of the correction function $X(\lambda)$. Comparison of two
lines in (\ref{Psi_as_s1+s2ne0}) yields the asymptotics of the Painlev\'e 
function, $y=y_1+2\hat h_1$, which turns into (\ref{mcleod-_as}) for 
$\Re(\sigma\alpha)\geq0$, $\alpha-\frac{1}{2}\notin{\Bbb Z}$ substituting 
$\nu=\sigma\alpha/2$.

Validity of the asymptotic formula (\ref{mcleod-_as}) without the restriction 
(\ref{nu>=0}) follows from the observation that (\ref{mcleod-_as}) is 
invariant with respect to B\"acklund transformations. Indeed, the
Schlesinger transformation of the $\Psi$-function,
\begin{multline}\label{Schlesinger}
\tilde\Psi(\lambda)=R(\lambda)\Psi(\lambda),
\\
R(\lambda)=I
-\frac{q_{\epsilon}}{2i\lambda}(\sigma_3-i\epsilon\sigma_1),\quad
q_{\epsilon}=
\frac{\alpha+\frac{\epsilon}{2}}{y_x-\epsilon(y^2+\frac{x}{2})},\quad
\epsilon\in\{+1,-1\},
\end{multline}
yields the $\Psi$-function, associated to a new Painlev\'e transcendent
\begin{equation}\label{Backlund}
\tilde y=y+\epsilon q_{\epsilon},\quad
\tilde y_x=y_x+\epsilon(\tilde y^2-y^2),\quad
\tilde\alpha=-\alpha-\epsilon,
\end{equation}
characterized however by the same Stokes multipliers. The latter with the
definition of the parameter $\sigma$ via
$s_1=e^{-i\pi\sigma(\alpha+\frac{1}{2})}=
\tilde s_1=e^{-i\pi\tilde\sigma(\tilde\alpha+\frac{1}{2})}$ implies the
transformation of the parameter $\sigma$, $\tilde\sigma=-\sigma$ for either
value of $\epsilon=\pm1$. If both $\Re(\sigma\alpha)\geq0$ and
$\Re(\tilde\sigma\tilde\alpha)=\Re(\sigma\alpha)+\sigma\epsilon\geq0$, then
the asymptotics of both $y$ and $\tilde y$ is described by (\ref{mcleod-_as})
(supplemented by tilde over $y$, $\sigma$ and $\alpha$ if necessary). This 
observation constitutes the invariance of the asymptotic formula with respect 
to the B\"acklund transformation. Since the latter transformation is a 
bi-rational transformation in the space with coordinates $(y,y_x)$, the above 
invariance can be confirmed algebraically. Therefore, if 
$\Re(\tilde\sigma\tilde\alpha)<0$ while $\Re(\sigma\alpha)\geq0$, the
asymptotics of $\tilde y$ and $y$ are described by the formula 
(\ref{mcleod-_as}) (supplemented by tilde if necessary). The iterated use 
of the B\"acklund transformations completes the proof.
\endproof

\subsection{The increasing degenerate Painlev\'e functions}

Applying the second of the symmetries (\ref{P_symmetries}) to
(\ref{mcleod-_as}) and changing the argument of $x$ in $2\pi$, we obtain

\begin{theorem}\label{McLeod3}
If $\alpha-\frac{1}{2}\notin{\Bbb Z}$, $1+s_0s_1=0$, 
$\arg x\in[\pi,\frac{4\pi}{3}]$ and $|x|$ is large enough, 
then the asymptotics of the Painlev\'e function is given by
\begin{multline}\label{mcleod-bar_as}
y=y_0(x,\alpha,\sigma)-
\\
-\frac{s_2-s_0}{\pi}
2^{-\frac{5}{2}\sigma\alpha-\frac{7}{4}}
\Gamma(\tfrac{1}{2}+\sigma\alpha)
(e^{-i\pi}x)^{-\frac{3}{2}\sigma\alpha-\frac{1}{4}}
e^{-\frac{2\sqrt2}{3}(e^{-i\pi}x)^{3/2}}
\bigl(1+{\cal O}(x^{-1/4})\bigr),
\end{multline}
where $y_0(x,\alpha,\sigma)=\overline{y_1(\bar x,\bar\alpha,\sigma)}
\sim\sigma\sqrt{e^{-i\pi}x/2}$ is the solution of the Painlev\'e equation
corresponding to $1+s_0s_1=s_2-s_0=0$, while the parameter 
$\sigma\in\{+1,-1\}$ is defined by the use of the equation 
$s_1=e^{-i\pi\sigma(\alpha+\frac{1}{2})}$.
\end{theorem}

The solutions $y_1(x,\alpha,\sigma)$ and 
$y_0(x,\alpha,\sigma)=\overline{y_1(\bar x,\bar\alpha,\sigma)}$ are meromorphic 
functions of $x\in{\Bbb C}$ and thus can be continued beyond the sectors 
indicated in Theorems~\ref{McLeod2} and~\ref{McLeod3}. To find the asymptotics 
of the solution $y_0(x,\alpha,\sigma)$ in the interior of the sector 
$\arg x\in[\frac{2\pi}{3},\pi]$, we apply (\ref{mcleod-_as}) with $s_2=s_0$. 
Similarly, we find the asymptotics of the solution $y_1(x,\alpha,\sigma)$ in 
the interior of the sector $\arg x\in[\pi,\frac{4\pi}{3}]$ using 
(\ref{mcleod-bar_as}) with $s_2=-s_1$. Either expression implies that, if 
$|x|\to\infty$, $\arg x\in[\frac{2\pi}{3},\frac{4\pi}{3}]$,
\begin{multline}\label{y0-y1}
y_0(x,\alpha,\sigma)-y_1(x,\alpha,\sigma)=
\\
=i\sigma
\frac{2^{-\frac{5}{2}\sigma\alpha-\frac{3}{4}}}
{\Gamma(\tfrac{1}{2}-\sigma\alpha)}
(e^{-i\pi}x)^{-\frac{3}{2}\sigma\alpha-\frac{1}{4}}
e^{-\frac{2\sqrt2}{3}(e^{-i\pi}x)^{3/2}}
\bigl(1+{\cal O}(x^{-1/4})\bigr),
\end{multline}
where we have used the relation
$$
(s_0+s_1)\Gamma\bigl(\tfrac{1}{2}+\sigma\alpha\bigr)=
-\frac{2i\pi\sigma}{\Gamma\bigl(\tfrac{1}{2}-\sigma\alpha\bigr)}.
$$
The formula (\ref{y0-y1}) constitutes the quasi-linear Stokes phenomenon for
the increasing degenerate asymptotic solution of $P_2$.

Due to exponential decay of the difference (\ref{y0-y1}) in the interior of
the sector $\arg x\in\bigl(\frac{2\pi}{3},\frac{4\pi}{3}\bigr)$, we have
solutions of $P_2$,
\begin{multline}\label{y0_ext1}
y=y_1(x,\alpha,\sigma)\simeq\sigma\sqrt{e^{-i\pi}x/2},\quad
|x|\to\infty,\quad
\arg x\in\bigl[\tfrac{2\pi}{3},\tfrac{4\pi}{3}\bigr),
\\
\text{for}\quad
s_0=-e^{i\pi\sigma(\alpha+\frac{1}{2})},\quad
s_1=e^{-i\pi\sigma(\alpha+\frac{1}{2})},\quad
s_2=-e^{-i\pi\sigma(\alpha+\frac{1}{2})},
\end{multline}
and
\begin{multline}\label{y1_ext0}
y=y_0(x,\alpha,\sigma)\simeq\sigma\sqrt{e^{-i\pi}x/2},\quad
|x|\to\infty,\quad
\arg x\in\bigl(\tfrac{2\pi}{3},\tfrac{4\pi}{3}\bigr],
\\
\text{for}\quad
s_0=-e^{i\pi\sigma(\alpha+\frac{1}{2})},\quad
s_1=e^{-i\pi\sigma(\alpha+\frac{1}{2})},\quad
s_2=-e^{i\pi\sigma(\alpha+\frac{1}{2})}.
\end{multline}

Applying the rotational symmetry of (\ref{P_symmetries}) to $y_1$ and
$y_0$, we find solutions
\begin{multline}\label{y_2n+1_def}
y=y_{2n+1}(x,\alpha,\sigma):=
e^{i\frac{2\pi}{3}n}y_1(e^{i\frac{2\pi}{3}n}x,\alpha,\sigma)\simeq
\sigma(-1)^n\sqrt{e^{-i\pi}x/2},
\\
|x|\to\infty,\quad
\arg x\in\bigl[\tfrac{2\pi}{3}-\tfrac{2\pi}{3}n,
\tfrac{4\pi}{3}-\tfrac{2\pi}{3}n\bigr),
\\
\text{for}\quad
s_{2n}=-e^{i\pi\sigma(\alpha+\frac{1}{2})},\quad
s_{2n+1}=e^{-i\pi\sigma(\alpha+\frac{1}{2})},\quad
s_{2n+2}=-e^{-i\pi\sigma(\alpha+\frac{1}{2})},
\end{multline}
and 
\begin{multline}\label{y_2n_def}
y=y_{2n}(x,\alpha,\sigma):=
e^{i\frac{2\pi}{3}n}y_0(e^{i\frac{2\pi}{3}n}x,\alpha,\sigma)\simeq
\sigma(-1)^n\sqrt{e^{-i\pi}x/2},
\\
|x|\to\infty,\quad
\arg x\in\bigl(\tfrac{2\pi}{3}-\tfrac{2\pi}{3}n,
\tfrac{4\pi}{3}-\tfrac{2\pi}{3}n\bigr],
\\
\text{for}\quad
s_{2n}=-e^{i\pi\sigma(\alpha+\frac{1}{2})},\quad
s_{2n+1}=e^{-i\pi\sigma(\alpha+\frac{1}{2})},\quad
s_{2n+2}=-e^{i\pi\sigma(\alpha+\frac{1}{2})}.
\end{multline}

Comparing the Stokes multipliers, we observe the symmetries
\begin{equation}\label{y_n_symm}
y_n(x,\alpha,\sigma)=y_{n+6}(x,\alpha,\sigma),\quad
y_{2n-1}(x,\alpha,\sigma)=y_{2n}(x,\alpha,-\sigma).
\end{equation}
Equation (\ref{y0-y1}) and definitions (\ref{y_2n+1_def}) and
(\ref{y_2n_def}) imply the differences take place,
\begin{multline}\label{y_2n+1-y_2n}
y_{2n}(x,\alpha,\sigma)-y_{2n+1}(x,\alpha,\sigma)=
\\
=i\sigma
\frac{2^{-\frac{5}{2}\sigma\alpha-\frac{3}{4}}}
{\Gamma(\tfrac{1}{2}-\sigma\alpha)}
e^{i\frac{2\pi}{3}n}
(e^{-i\pi}e^{i\frac{2\pi}{3}n}x)^{-\frac{3}{2}\sigma\alpha-\frac{1}{4}}
e^{-\frac{2\sqrt2}{3}(e^{-i\pi}e^{i\frac{2\pi}{3}n}x)^{3/2}}
\bigl(1+{\cal O}(x^{-1/4})\bigr),
\\
|x|\to\infty,\quad
\arg
x\in\bigl[\tfrac{2\pi}{3}-\tfrac{2\pi}{3}n,
\tfrac{4\pi}{3}-\tfrac{2\pi}{3}n\bigr],
\end{multline}
which are exponentially small in the interior of the indicated sectors.

Using the second of the equations (\ref{y_n_symm}), we establish the
analytic continuation of the asymptotics as $|x|\to\infty$,
\begin{multline}\label{max_as_sect}
y_{2n-1}(x,\alpha,(-1)^{n-1}\sigma)=y_{2n}(x,\alpha,(-1)^n\sigma)\simeq
\sigma\sqrt{e^{-i\pi}x/2},
\\
|x|\to\infty,\quad
\arg x\in\bigl(\tfrac{2\pi}{3}(1-n),\tfrac{2\pi}{3}(3-n)\bigr).
\end{multline}

\section{Coefficient asymptotics for $\alpha-\frac{1}{2}\not\in{\Bbb Z}$}

An elementary investigation shows that the equation $P_2$,
$y_{xx}=2y^2+xy-\alpha$, can be satisfied by the formal series depending 
on $\alpha$ and a parameter $\sigma=\pm1$,
\begin{equation}\label{mcleod-formal}
y_f(x,\alpha,\sigma)=
\sigma\sqrt{-x/2}\sum_{n=0}^{\infty}b_n(-x)^{-3n/2}
+{\cal O}(x^{-\infty}).
\end{equation}
Given $\sigma$, the series (\ref{mcleod-formal}) is determined uniquely
since its coefficients $b_n$ are determined by the recurrence relation,
\begin{multline}\label{bn_recurrence}
b_0=1,\quad 
b_1=\frac{\sigma\alpha}{\sqrt2},
\\
b_{n+2}=\frac{9n^2-1}{8}b_{n}
-\sum_{m=1}^{n+1}b_mb_{n+2-m}
-\frac{1}{2}\sum_{l=1}^{n+1}\sum_{m=1}^{n+2-l}b_lb_mb_{n+2-l-m}.
\end{multline}
Several initial terms of the expansion are given by
\begin{multline}\label{mcleod_expansion}
y_f(x,\alpha,\sigma)=\sigma\sqrt{-x/2}\Bigl\{1+
\frac{\sigma\alpha}{\sqrt2(-x)^{3/2}}
-\frac{1+6\alpha^2}{8(-x)^{3}}
+\frac{\sigma\alpha(11+16\alpha^2)}{8\sqrt2(-x)^{9/2}}-
\\
-\frac{73+708\alpha^2+420\alpha^4}{128(-x)^{6}}
+\frac{\sigma\alpha(1021+2504\alpha^2+768\alpha^4)}
{64\sqrt2(-x)^{15/2}}-
\\
-\frac{10657+129918\alpha^2+132060\alpha^4+24024\alpha^6}
{1024(-x)^{9}}+
\\
+\frac{\sigma\alpha(248831+786304\alpha^2+416400\alpha^4+49152\alpha^6)}
{512\sqrt2(-x)^{21/2}}
+{\cal O}\bigl(x^{-12}\bigr)\Bigr\}.
\end{multline}

To find the asymptotics of the coefficients $b_n$ in (\ref{mcleod-formal}) 
as $n\to\infty$, let us construct a sectorial analytic function $\hat y(t)$,
\begin{multline}\label{hat_mcleod}
\hat y(t)=y_{2n-1}(e^{i\pi}t^2,\alpha,(-1)^{n-1}\sigma)
=y_{2n}(e^{i\pi}t^2,\alpha,(-1)^n\sigma),
\\
|t|\to\infty,\quad
\arg t\in\bigl(-\tfrac{\pi}{3}n,
\tfrac{\pi}{3}-\tfrac{\pi}{3}n\bigr).
\end{multline}
The function $\hat y(x)$ has a finite number of simple poles and, by
construction, has the uniform asymptotics near infinity, 
$\hat y(t)\simeq\sigma t/\sqrt2$ as $|t|\to\infty$. Using
(\ref{mcleod-formal}), it has the following formal series expansion,
\begin{equation}\label{mcleod_hat_y_expansion}
\hat y(t)=
\sigma\frac{t}{\sqrt2}\sum_{n=0}^{\infty}b_nt^{-3n}.
\end{equation}

Let $y^{(N)}(x)$ be a partial sum
\begin{equation}\label{mcleod_yN}
y^{(N)}(t)=\sigma\frac{t}{\sqrt2}\sum_{n=0}^{N-1}b_nt^{-3n},
\end{equation}
and $v^{(N)}(t)$ be a product
\begin{equation}\label{mcleod_vn}
v^{(N)}(t)=t^{3N-2}(\hat y(t)-y^{(N)}(t))=
\frac{\sigma}{t\sqrt2}\sum_{n=0}^{\infty}b_{n+N}t^{-3n}.
\end{equation}
Because $t^{3N-2}y^{(N)}(t)$ is polynomial and $\hat y(t)/t$ is bounded as
$|t|\geq\rho$, the integral of $v^{(N)}(t)$ along the circle of the radius 
$|t|=\rho$ containing all the pole singularities of $\hat y(t)$ satisfies 
the estimate
\begin{equation}\label{mcleod_vN_integral}
\Bigl|\oint_{|t|=\rho}v^{(N)}(t)\,dt\Bigr|\leq
\rho^{3N-2}\oint_{|t|=\rho}|\hat y(t)|\,dl\leq
2\pi\rho^{3N}\max_{|t|=\rho}|\hat y(t)/t|=C\rho^{3N}.
\end{equation}
On the other hand, inflating the sectorial arcs of the circle
$|t|=\rho$, we find that
\begin{equation}\label{mcleod_vN_inflation}
\oint_{|t|=\rho}v^{(N)}(t)\,dx=\oint_{|t|=R}v^{(N)}(t)\,dt
+\sum_{n=-3}^2\int_{e^{i\frac{\pi}{3}n}(\rho,R)}\bigl(
v_+^{(N)}(t)-v_-^{(N)}(t)\bigr)\,dt.
\end{equation}

Since $v^{(N)}(t)=\frac{\sigma}{t\sqrt2}b_N+{\cal O}(t^{-3})$, the first
of the integrals in the r.h.s.\ of (\ref{mcleod_vN_inflation}) is
\begin{equation}\label{bN_int}
\oint_{|x|=R}v^{(N)}(t)\,dt=\pi i\sigma\sqrt{2}\,b_N+{\cal O}(R^{-2}).
\end{equation}
Last six integrals in (\ref{mcleod_vN_inflation}) are computed using
(\ref{hat_mcleod}), (\ref{y_2n+1_def}), (\ref{y_2n_def}), (\ref{y_n_symm})
and (\ref{y0-y1}),
\begin{multline}\label{bN_eval}
\sum_{n=-3}^2\int_{e^{i\frac{\pi}{3}n}(\rho,R)}\bigl(
v_+^{(N)}(t)-v_-^{(N)}(t)\bigr)\,dt=
\\
=3\int_{\rho}^R t^{3N-2}\bigl(
y_0(e^{i\pi}t^2,\alpha,\sigma)
-y_1(e^{i\pi}t^2,\alpha,\sigma)\bigr)\,dt+
\\
+3(-1)^{N-1}\int_{\rho}^R t^{3N-2}\bigl(
y_0(e^{i\pi}t^2,\alpha,-\sigma)
-y_1(e^{i\pi}t^2,\alpha,-\sigma)\bigr)\,dt=
\\
=3i\sigma
\frac{2^{-\frac{5}{2}\sigma\alpha-\frac{3}{4}}}
{\Gamma(\tfrac{1}{2}-\sigma\alpha)}
\int_{\rho}^R
t^{3(N-\sigma\alpha)-\frac{5}{2}}
e^{-\frac{2\sqrt2}{3}t^3}
\bigl(1+{\cal O}(t^{-1/2})\bigr)\,dt+
\\
+3(-1)^{N}i\sigma
\frac{2^{\frac{5}{2}\sigma\alpha-\frac{3}{4}}}
{\Gamma(\tfrac{1}{2}+\sigma\alpha)}
\int_{\rho}^R 
t^{3(N+\sigma\alpha)-\frac{5}{2}}
e^{-\frac{2\sqrt2}{3}t^3}
\bigl(1+{\cal O}(t^{-1/2})\bigr)\,dt=
\\
=i\sigma
2^{-\frac{3}{2}N}
3^{N-\frac{1}{2}}
\Bigl\{
6^{-\sigma\alpha}
\frac{\Gamma(N-\sigma\alpha-\tfrac{1}{2})}
{\Gamma(\tfrac{1}{2}-\sigma\alpha)}
\bigl(1+{\cal O}(N^{-1/6})\bigr)+
\\
+(-1)^N
6^{\sigma\alpha}
\frac{\Gamma(N+\sigma\alpha-\tfrac{1}{2})}
{\Gamma(\tfrac{1}{2}+\sigma\alpha)}
\bigl(1+{\cal O}(N^{-1/6})\bigr)
\Bigr\}
+
\\
+{\cal O}(\rho^{3(N+|\mbox{\tiny\rm Re}\alpha|)-\frac{3}{2}})
+{\cal O}\bigl(e^{-\frac{2\sqrt2}{3}R^3}
R^{3(N+|\mbox{\tiny\rm Re}\alpha|)-\frac{9}{2}}\bigr).
\end{multline}
Thus, letting $R=\infty$, we find the asymptotics as $N\to\infty$ of
the coefficient $b_N$ in (\ref{mcleod-formal}),
\begin{multline}\label{bn_fin}
b_N=-\frac{1}{\pi\sqrt{6}}
\Bigl(\frac{3}{2\sqrt2}\Bigr)^N
\Bigl\{
6^{-\sigma\alpha}
\frac{\Gamma(N-\sigma\alpha-\tfrac{1}{2})}
{\Gamma(\tfrac{1}{2}-\sigma\alpha)}
\bigl(1+{\cal O}(N^{-1/6})\bigr)+
\\
+(-1)^N
6^{\sigma\alpha}
\frac{\Gamma(N+\sigma\alpha-\tfrac{1}{2})}
{\Gamma(\tfrac{1}{2}+\sigma\alpha)}
\bigl(1+{\cal O}(N^{-1/6})\bigr)
\Bigr\}
+{\cal O}(\rho^{3N}).
\end{multline}
In particular, the asymptotics (\ref{bn_fin}) is consistent with the
elementary observation that, for $\alpha=0$, all odd coefficients vanish, 
$b_{2n-1}=0$; for even coefficients, the leading order asymptotics reduces 
to the one-term expression, i.e.
\begin{multline}\label{bn_fin_a=0}
\text{if}\quad
\alpha=0\colon\quad
b_{2N-1}=0,
\\
b_{2N}=-\frac{\sqrt2}{\pi^{3/2}\sqrt{3}}
\Bigl(\frac{3}{2\sqrt2}\Bigr)^{2N}
\Gamma(2N-\tfrac{1}{2})
\bigl(1+{\cal O}(N^{-1/6})\bigr)
+{\cal O}(\rho^{3N}).
\end{multline}
\begin{rem}\label{mcleod_err_bound}
The error bound in (\ref{mcleod-_as}), (\ref{mcleod-bar_as}) and
(\ref{y0-y1}) can be improved to ${\cal O}(x^{-3/2})$ which implies the
error bound for (\ref{bn_fin}) ${\cal O}(N^{-1})$.
\end{rem}

\begin{rem}\label{alpha=1/2+n}
It is possible to prove that the asymptotic formula (\ref{bn_fin}) remains
valid for half-integer values of $\alpha$ as well.
\end{rem}

\bigskip
{\bf Acknowledgments.} The author was supported in part by 
the RFBR, grant No.~02--01--00268. He also thanks Prof.~A.R.~Its for
stimulating discussions.

\ifx\undefined\bysame \newcommand{\bysame}{\leavevmode\hbox
to3em{\hrulefill}\,} \fi

\end{document}